\documentclass[a4paper,11pt]{article}
\usepackage{pos}
\usepackage{graphics}
\usepackage{siunitx}
\usepackage{booktabs}
\usepackage{setspace}

\DeclareSIUnit{\sample}{Sa}
\DeclareSIUnit{\pe}{p.e.}

\title{Detection of the Crab Nebula by the prototype Schwarzschild-Couder Telescope}
 \ShortTitle{pSCT Detection of the Crab Nebula}

\author*[a]{B.~A.~W.~Mode}
\affiliation[a]{University of Wisconsin - Madison,\\
  Department of Physics and Wisconsin IceCube Particle Astrophysics Center, 222 W Washington Ave, Madison, WI, USA}

\forColl{CTA SCT}
\emailAdd{bmode@wisc.edu}
\abstract{The Schwarzschild-Couder Telescope (SCT) is a medium-sized telescope technology proposed for the Cherenkov Telescope Array. It uses a novel dual-mirror optical design that removes comatic aberrations across its entire field of view. The SCT camera employs high-resolution silicon photomultiplier (SiPM) sensors with a pixel size of 4 arcminutes. A prototype SCT (pSCT) has been constructed at the Fred Lawrence Whipple Observatory in Arizona, USA. An observing campaign in 2020, with a partial camera of 1600 pixels (2.7 degrees by 2.7 degrees field of view)  resulted in detection of the Crab Nebula at 8.6 sigma statistical significance. Work on the pSCT camera and optical system is ongoing to improve performance and prepare for an upcoming camera upgrade. The pSCT camera upgrade will replace the current camera modules with improved SiPMs and readout electronics and will expand the camera to its full design field of view of 8 degrees in diameter (11,328 pixels). The fully upgraded pSCT will enable next-generation very-high-energy gamma-ray astrophysics through excellent background rejection and angular resolution. In this presentation we describe first results from the successful operation of the pSCT and future plans.}

\FullConference{37$^{\rm{th}}$ International Cosmic Ray Conference (ICRC 2021)\\
		July 12th -- 23rd, 2021\\
		Online -- Berlin, Germany}

\begin{document}
\maketitle

\section{Introduction}

The multi-messenger paradigm has come to hold an important position within astrophysics, combining observations from different instruments to better understand astrophysical phenomena. What has enabled multi-messenger analyses are multiple recent advances in observing technologies that go beyond the traditional messengers and energy scales of previous astronomical efforts. The development of neutrino observatories such as IceCube and gravitational wave observatories has opened new modalities. Of the relatively recent advances that allow multi-messenger astrophysics, one of the most central has been the observation of astrophysical very-high-energy (VHE) gamma rays. There are three complementary methods for detecting astrophysical VHE gamma rays. The first was satellite-based observatories such as the Compton Gamma-Ray Observatory or Fermi-LAT. This approach is limited to the hard x-ray through high-energy gamma ray portion of the electromagnetic spectrum, and is further limited by its small effective area. The effective areas of experiments using either of two terrestrial detection methods are substantially larger than that which is feasible for satellite-based observatories.

Another approach employed by LHAASO \cite{lhaaso} (as a component), HAWC \cite{hawc}, and soon SWGO \cite{swgo}, is to provide a water Cherenkov target inside of tanks instrumented with photosensors. The VHE gamma initiates an air shower in the atmosphere, and the shower particles interact with the water target, including electrically charged particles that emit Cherenkov radiation. This technique is valid at much higher energies and benefits from an extremely wide field of view (FOV); however, the angular resolution and background rejection of such instruments is poor. Complementary to the water target is the imaging atmospheric Cherenkov telescope (IACT). VHE gamma-initiated air showers include charged particles whose velocity in the atmosphere exceeds the speed of light in the same medium. These charged particles excite the dielectric medium, and the photons emitted constitute Cherenkov radiation. The resulting cone of radiation can be captured by a telescope with a sufficiently fast photosensing camera. The IACT approach has a much narrower field of view than the water Cherenkov observatories, but offers much greater angular resolution in return. It is also possible to construct IACTs that are sensitive over a wide swathe of the VHE gamma-ray spectrum in a manner that is complementary in energy range to the other techniques for observing gamma rays. Note that LHAASO has seen recent success in ultra-high-energy gamma-ray physics, detecting signals from 12 potential Galactic PeVatrons \cite{lhaaso_pev}.

The first IACT was the Whipple 10 m telescope at the Fred Lawrence Whipple Observatory (FLWO) south of Tucson, AZ, USA. This telescope used single-mirror Davies-Cotton optics and began its life with a camera instrumented with photomultiplier tubes (PMTs) tuned for visible light \cite{whipple10m}. The night sky background light proved to be unnecessarily prominent; after a few years of regular operation, the camera was reinstrumented with PMTs designed for the UV spectrum \cite{whippleuv}. The design choices of tessellated Davies-Cotton mirror panels and UV-sensitive PMT-based cameras became the standard for IACTs.

The Cherenkov Telescope Array (CTA) is the next-generation gamma-ray observatory. It will consist of a southern hemisphere array located in Paranal, Chile on European Southern Observatory grounds, and a northern hemisphere array located on the island of La Palma in the Canary Islands, Spain, near MAGIC, an existing IACT array. The southern array is planned to be sensitive over the full CTA energy range from 20 GeV to 300 TeV. The northern array will be sensitive from 20 GeV to 50 TeV. In order to be sensitive across this energy range, CTA will utilize a combination of different telescope designs that each centers on a particular subregion of the target energy sensitivity range. 

As part of CTA, we are developing a medium-size telescope candidate that uses a novel dual-mirror optical design and a compact, high-resolution camera. We describe our observations of the Crab Nebula resulting in the first source detection using our prototype telescope. While this detection was made using the current prototype camera, an ongoing upgrade to the camera and other subsystems will greatly increase the FOV and lower the energy threshold of the instrument. The detection of the Crab Nebula is described in more depth in our recently published article \cite{crab}.

\section{The prototype Schwarzschild-Couder Telescope}

The Schwarzschild-Couder Telescope (SCT) is a candidate design for a medium-sized telescope for CTA. Unlike prior IACT designs, the SCT uses a dual-mirror Schwarzschild-Couder optics to improve off-axis performance over Davies-Cotton (DC) optics and a camera instrumented with silicon photomultipliers (SiPMs). The SCT offers several advantages over traditional DC optics in terms of a much improved off-axis point spread function, a wider field of view, and a smaller plate scale \cite{vladimir}. The smaller plate scale motivates the use of SiPMs in the SCT camera in order to achieve image resolution comparable to the optical point-spread function (0.067 square degrees per pixel). 

A prototype SCT (pSCT) has been constructed at FLWO, at the same site as the VERITAS IACT array. The pSCT sits 35 m from the closest VERITAS telescope, T4. The pSCT camera is currently only instrumented in its central sector, with 1600 image pixels and a \SI{2.7}{\degree} by \SI{2.7}{\degree} square FOV. The current camera consists of 25 focal plane modules using two different SiPM models. Nine of the camera modules use SiPMs produced by FBK, and the remaining modules use SiPMs produced by Hamamatsu. Triggering and digitization are provided by TARGET 7 (7th generation TeV Array Readout with \SI{}{\giga\sample\per\second} sampling and Event Trigger) application-specific integrated circuits (ASICs). A feature of TARGET is a \SI{16}{\micro\second} long storage array to allow for cross-telescope array triggers as part of CTA. An upgrade of several critical systems including fully populating the camera with updated SiPMs and front-end electronics is underway for the pSCT. This will be described in more detail in a later section.

\section{Observation of the Crab Nebula}

Initial pSCT observations of the Crab Nebula ran from January 18, 2020 to February 26, 2020. Data were taken primarily in \SI{60}{\minute} runs in an on-source/off-source observing mode. A typical run consisted of \SI{29}{\minute} of on-source pointing, \SI{29}{\minute} of off-source pointing, and \SI{2}{min} to slew between the two positions. The ordering of on- and off-source pointing was determined to be whichever order maximized the elevation of on-source pointing. Off-source pointing consisted of tracking the same patch of sky relative to Earth, i.e. an appropriate half hour offset in right ascension. Data taking typically occurred at a rate of \SI{100}{\hertz} with an additional \SI{10}{\hertz} of injected LED flasher pulses for calibration purposes. The total livetime was \SI{17.6}{\hour} off-source and \SI{21.6}{\hour} on-source. Only on-source data with matching off-source data was used to calculate detection significance; the remaining four hours of on-source data was used for cut selection.

\begin{figure}[ht]
    \centering
    \includegraphics[width=\textwidth]{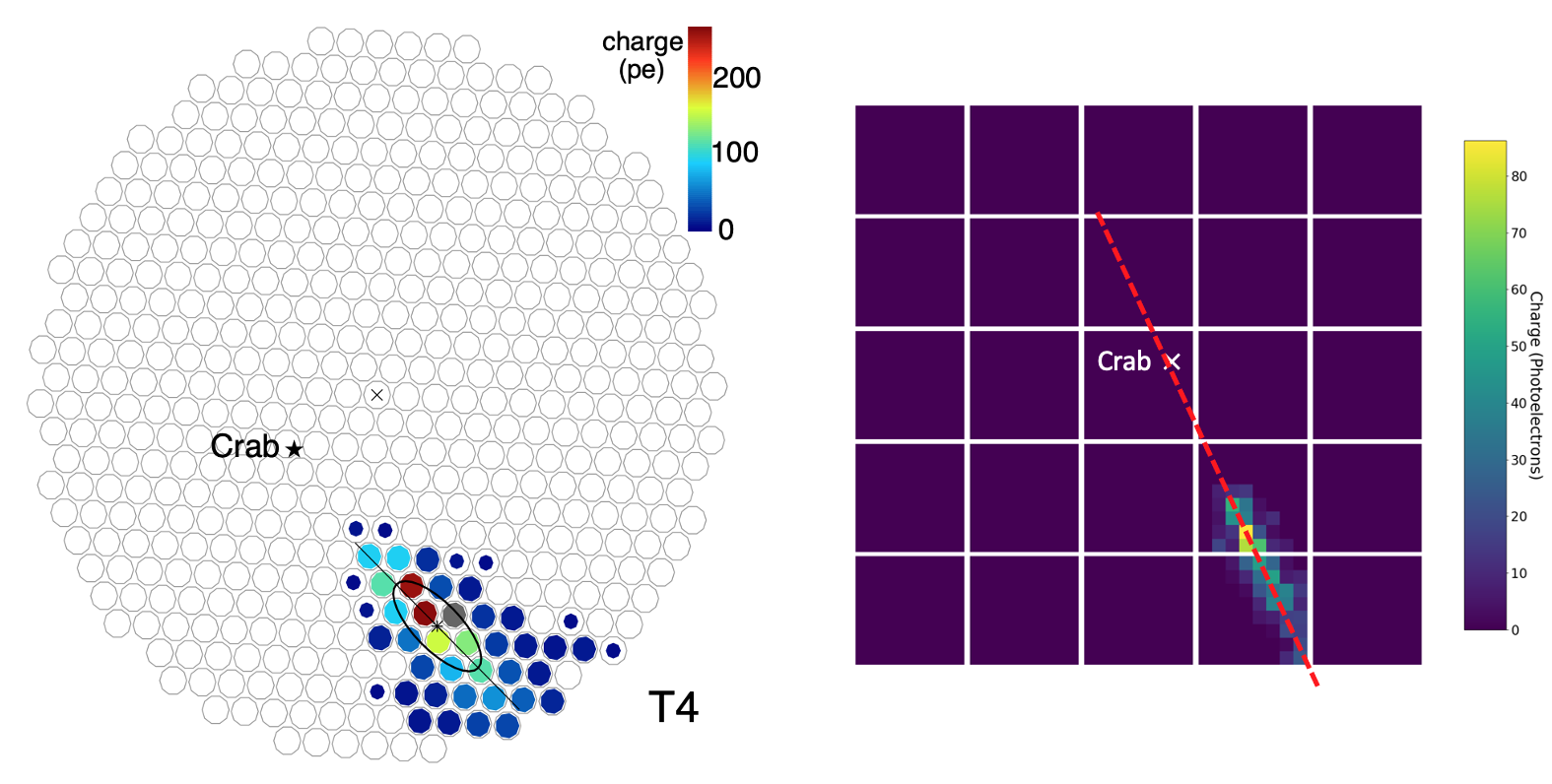}
    \captionof{figure}{Coincident event between VERITAS telescope T4 (left) and pSCT (right). This event was identified between data sets using event timing. It is qualitatively clear that both telescopes have recorded the same Cherenkov shower from a similar viewing angle.}
    \label{ref:dualevent}
\end{figure}

In order to distinguish between gamma-initiated showers and hadronic showers, the pSCT and VERITAS observed the Crab Nebula in coincidence. For an example of event coincidence between VERITAS and pSCT, see Figure \ref{ref:dualevent}. While this technique is less standard than simulations, using the well-established VERITAS analysis as the basis to calibrate our analysis is more powerful. Most novel instruments are not co-located near well-established observing arrays, which limits the wide-spread application of such a technique. This included a set of three \SI{60}{\minute} observing runs on-source entirely coincident with VERITAS. In our analysis, we identify events triggered coincidentally by an analysis of event timing. This resulted in a calibration sample of 18 coincident gamma-identified showers and 11,597 hadronic-identified showers, where these categorizations come from the standard VERITAS analysis of their event data. 

\section{Initial Analysis Using VERITAS Coincidence}

On the entire selection of Crab Nebula data, we performed a moment analysis inspired by \cite{whipple10m} and \cite{whippleuv}. In particular, our analysis made use of an aperture-based image cleaning procedure described by \cite{wood}, the cleaned image summed charge (size), and the geometric ellipse parameterization of \cite{hillas}. The aperture cleaning method involves calculating an image size based on the appropriately weighted aperture of radius 2 image pixels and performing a threshold cut based on noise levels. The resulting 2D images are parameterized into ellipse length and width, centroid, and resulting angles between the semi-major axis and the line from centroid to center of FOV ($\alpha$). 

A notable feature of the data set was that the Crab Nebula was not exactly centered in the FOV. The pointing for each event was corrected by using SiPM currents and known bright stars near the Crab Nebula within the FOV to derive a simple cubic polynomial fit correction model based on time relative to the culmination time of the Crab Nebula. This changes the calculation of the \textit{distance} Hillas parameter (distance from centroid to center of FOV) to the Hillas \textit{displacement} parameter (distance from centroid to pointing location), a more useful parameter for calculating the significance of observations and a significance sky map.

\begin{table}[ht]
    \centering
    \begin{tabular}{c} 
    \toprule
    Gamma-ray selection cuts \\  \midrule
    $0.33^{\circ} < distance < 1.14^{\circ}$\\
    $\SI{250}{\pe} < size < \SI{20000}{\pe}$\\
    $width < -0.070^{\circ}+0.047^{\circ}\log_{10}(size/p.e.)$ \\
    $length < -0.369^{\circ}+0.201^{\circ}\log_{10}(size/p.e.)$\\
    $1.139^{\circ}-1.742^{\circ}(width/length) < distance$\\
    $distance < 1.273^{\circ}-0.737^{\circ}(width/length)$\\
    \bottomrule
    \end{tabular}
    \captionof{table}{Gamma-ray selection cuts optimized using VERITAS matched events.}
    \label{ref:first_cuts}
\end{table}

The first set of cuts was derived manually by considering the distributions of Hillas parameters after some loose cuts to remove noise-triggered events. Care was taken to retain $\sim{95\%}$ of the gamma-ray sample coincident with VERITAS. The set of optimized cuts for the analysis with VERITAS coincidence can be read in Table \ref{ref:first_cuts}. We reiterate here that cuts were developed only on coincident data, and then applied blind to the independent data.

\begin{figure}[ht]
    \centering
    \includegraphics[width=0.75\textwidth]{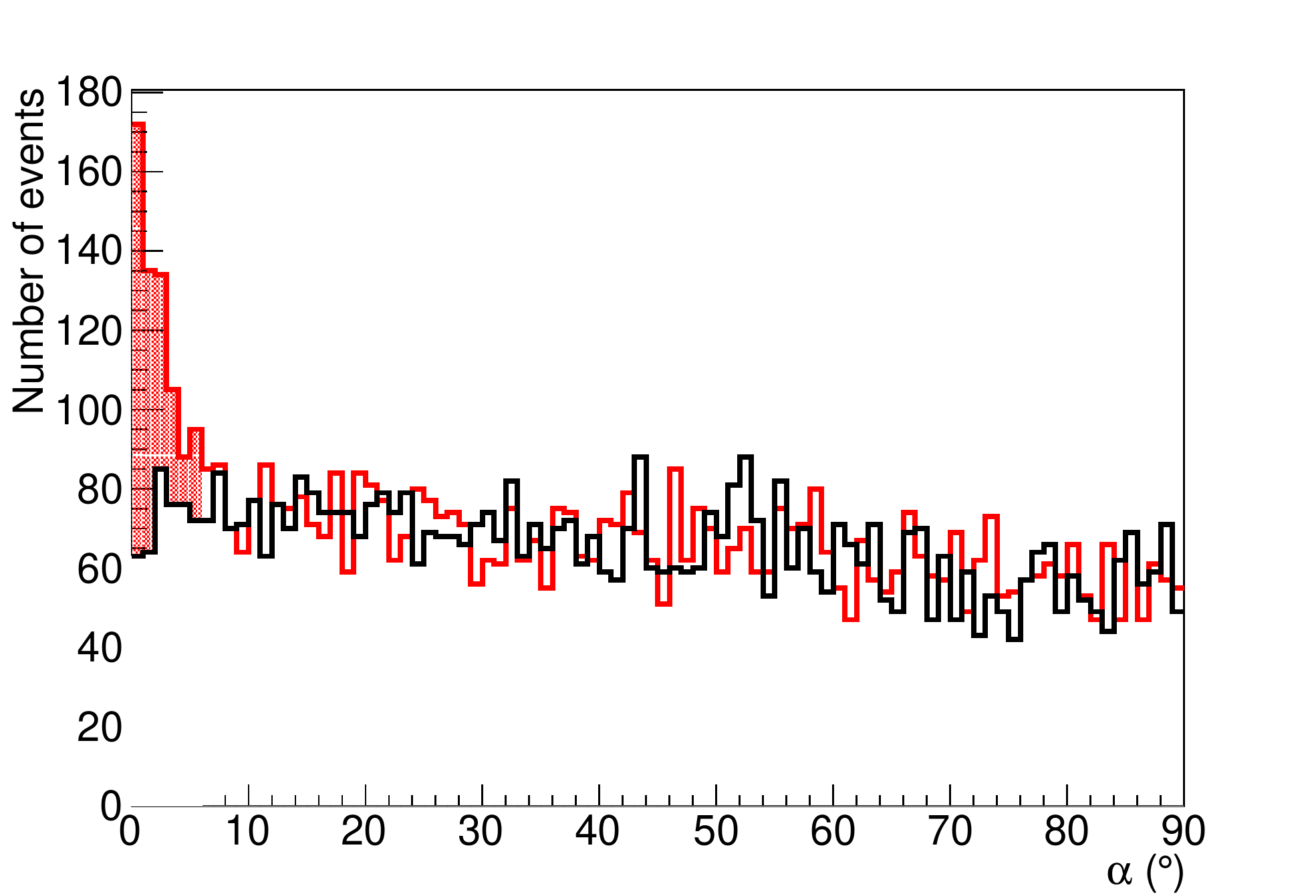}
    \captionof{figure}{Alpha (in degrees) is defined as the angle between the major axis of the shower image and the line joining the image centroid and source location in the field of view. The on-source distribution is represented in red, and the off-source distribution is represented in black. The observing duration was 17.6 hours for each. Each of these distributions is shown following the cuts introduced in Table \ref{ref:first_cuts}. Shaded is the region $\alpha < 6^{\circ}$.}
    \label{ref:alpha}
\end{figure}

The distribution of $\alpha$ following the preceding cuts can be seen in Figure \ref{ref:alpha}. Following the cut $\alpha < \SI{6}{\degree}$, we calculate using Equation 17 of \cite{li-ma} an excess of events in the on-source data of 8.6 $\sigma$ significance. The corresponding gamma-ray rate is ($0.28\pm 0.03$) \SI{}{\per\minute}. This rate is roughly consistent with expectations from our estimated energy threshold. Achieving a better understanding of the pSCT energy threshold and other calibration and sensitivity measurements are a main goal of the ongoing upgrade to the pSCT camera and subsystems.

\begin{figure}[ht]
    \centering
    \includegraphics[width=0.75\textwidth]{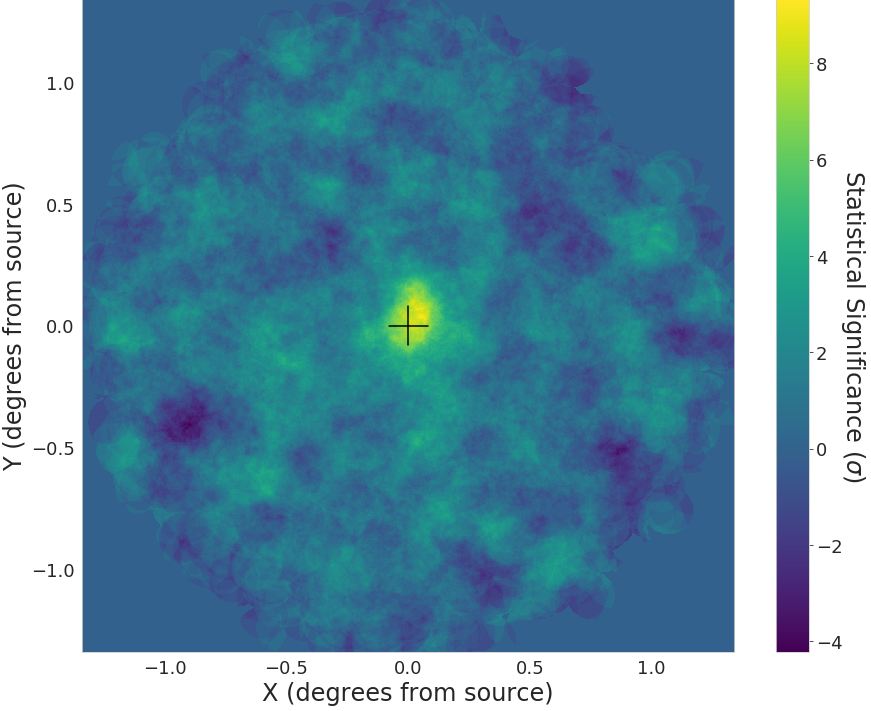}
    \captionof{figure}{Significance sky map in camera coordinates. The same event selection is used in both this sky map and Figure \ref{ref:alpha}. Shower arrival direction is determined using the method described in \cite{lessard}. A top-hat smoothing scheme of radius $0.1^{\circ}$ is applied, and then a Li-Ma significance calculated \cite{li-ma}. Here, X and Y are the azimuthal and elevation angles relative to the source in degrees.}
    \label{ref:skymap}
\end{figure}

We further illustrate the Crab Nebula detection with the pSCT by calculating a 2D sky map in camera coordinates using the method in \cite{lessard}. See Figure \ref{ref:skymap}. Note that this method relies on several Hillas parameters to estimate the arrival direction. Specifically, the arrival direction is estimated by considering the the angular distance along the line connecting the cleaned image centroid and the source position as

\begin{equation}
    disp = \xi(1 - \dfrac{width}{length})
\end{equation}

where $\xi$ is an empirically derived constant from the average of the \textit{distance} distribution for the VERITAS coincident gamma-ray sample. Specifically, for this sky map, $\xi=\SI{1.24}{\degree}$. The sky map is constructed by calculating Li-Ma statistical significances of each point on the map using the on- and off-source events estimated to fall within a \SI{0.1}{\degree} radius (top-hat smoothing). 

\section{Analysis Using pSCT Data Alone}

We were able to obtain a detection of the Crab Nebula above using data taken coincident with VERITAS. In order to explore improvements to the detection sensitivity of our analysis, we subdivided the \SI{17.6}{\hour} on- / off-source sample into a \SI{5.9}{\hour} ``training" sample and a \SI{11.7}{\hour} ``test" sample. Naively applying the cuts of the previous analysis to the training sample results in a 5.7 $\sigma$ significance detection, and applying the same cuts to the test sample results in a 6.7 $\sigma$ significance detection. We re-optimize the analysis by considering sequential changes to individual cut parameters and observing the effect on the significance of the detection when applied to the training sample. This approach favors simplicity; to achieve precisely optimized results, it would be necessary to perform a minimization over the full multidimensional parameter space. However, this procedure did improve pSCT detection performance over the initial analysis. This results in an updated set of cuts shown in Table \ref{ref:second_cuts}.

\begin{table}
    \centering
    \begin{tabular}{c} \toprule
    Gamma-ray selection cuts \\  \midrule
    $0.47^{\circ} < distance < 1.07^{\circ}$\\
    \SI{250}{\pe} $< size <$ \SI{20000}{\pe}\\
    $width < -0.072^{\circ}+0.045^{\circ}\log_{10}(size/p.e.)$ \\
    $length < -0.342^{\circ}+0.201^{\circ}\log_{10}(size/p.e.)$\\
    $\alpha < 6^{\circ}$\\
    \bottomrule
    \end{tabular}
    \captionof{table}{Gamma-ray selection cuts optimized using the pSCT training sample.}
    \label{ref:second_cuts}
\end{table}

Using this set of cuts, the \textit{a posteriori} significance of the training sample was 7.2 $\sigma$. Applying the new cuts \textit{a priori} to the test sample raises the significance from 6.7 $\sigma$ to 7.3 $\sigma$, a modest improvement. 

\section{pSCT Upgrade and Technical Improvements}

The CTA SCT Consortium is funded to upgrade the camera and several critical subsystems to achieve the design FOV and sensitivity using updated SiPM and electronics designs. The upgraded pSCT camera will be instrumented solely with an updated FBK SiPM model. The fully populated camera will have 11,328 imaging pixels to cover the entire design FOV of $\sim{8}$ \SI{}{\degree} diameter. This will require the production and validation of 177 camera modules. The front-end electronics will use updated TARGET ASICs for digitization (TARGET C) and triggering (T5TEA), splitting the functionality across two chips. An additional ASIC (SMART) is being designed and tested to provide robust pulse shaping of the raw SiPM output signal. The camera modules will connect to a control server via newly updated backplane and data acquisition (DACQ) boards. Backplanes implement the first stage of communication between camera modules and will communicate via the Distributed Intelligent Array Trigger (DIAT). The DIAT is also responsible for synchronizing the backplane clocks and generating a common event timestamp from GPS input. The DIAT system is also designed to allow multi-telescope trigger logic. The heat management system on the pSCT will also be upgraded with a more powerful design, capable of cooling the fully populated camera. More detailed information about the upgrade to the camera can be found in \cite{leslie}. Currently, the optical system is undergoing finer calibration as our alignment techniques improve \cite{deivid}.

\section{Acknowledgements}
This work was conducted in the context of the CTA SCT Collaboration.
We gratefully acknowledge financial support from the agencies and organizations listed here:
\url{http://www.cta-observatory.org/consortium}

\bibliographystyle{JHEP}
%\bibliography{references}
\bibliography{main.bbl}

\providecommand{\href}[2]{#2}\begingroup\raggedright\setstretch{0.01}\begin{thebibliography}{10}

\bibitem{lhaaso}
X.~Bai et~al., \emph{{The Large High Altitude Air Shower Observatory (LHAASO)
  Science White Paper}}, {\emph{arXiv e-print} (2019) arXiv:1905.02773}
  [\href{https://arxiv.org/abs/1905.02773}{{\ttfamily 1905.02773}}].

\bibitem{hawc}
A.~Smith, \emph{{HAWC: Design, Operation, Reconstruction, and analysis}},  in
  \emph{34th International Cosmic Ray Conference (ICRC34)}, International
  Cosmic Ray Conference, p.~996, July, 2015
  [\href{https://arxiv.org/abs/1508.05826}{{\ttfamily 1508.05826}}].

\bibitem{swgo}
U.B.~de~Almeida et~al., \emph{{The Southern Wide-Field Gamma-ray Observatory}},
  \href{https://doi.org/10.1002/asna.202113946}{\emph{Astron. Nachr.}
  {\bfseries 342} (2021) 431}.

\bibitem{lhaaso_pev}
Z.~Cao et~al., \emph{{Ultrahigh-energy photons up to 1.4 petaelectronvolts from
  12 $\gamma$-ray Galactic sources}},
  \href{https://doi.org/10.1038/s41586-021-03498-z}{\emph{Nature} {\bfseries
  594} (2021) 33}.

\bibitem{whipple10m}
T.C.~Weekes et~al., \emph{{Observation of TeV gamma rays from the Crab nebula
  using the atmospheric Cerenkov imaging technique}},
  \href{https://doi.org/10.1086/167599}{\emph{Astrophys. J.} {\bfseries 342}
  (1989) 379}.

\bibitem{whippleuv}
X.~Sarazin et~al., \emph{{Observation of the Crab Nebula with an ultraviolet
  Cherenkov camera}},
  \href{https://doi.org/10.1016/0927-6505(95)00042-9}{\emph{Astroparticle
  Physics} {\bfseries 4} (1996) 227}.

\bibitem{crab}
C.~Adams et~al., \emph{Detection of the crab nebula with the 9.7 m prototype
  schwarzschild-couder telescope},
  \href{https://doi.org/https://doi.org/10.1016/j.astropartphys.2021.102562}{\emph{Astroparticle
  Physics} {\bfseries 128} (2021) 102562}.

\bibitem{vladimir}
V.~Vassiliev et~al., \emph{{Wide field aplanatic two-mirror telescopes for
  ground-based {\ensuremath{\gamma}}-ray astronomy}},
  \href{https://doi.org/10.1016/j.astropartphys.2007.04.002}{\emph{Astroparticle
  Physics} {\bfseries 28} (2007) 10}.

\bibitem{wood}
M.~Wood et~al., \emph{Monte carlo studies of medium-size telescope designs for
  the cherenkov telescope array},
  \href{https://doi.org/https://doi.org/10.1016/j.astropartphys.2015.04.008}{\emph{Astroparticle
  Physics} {\bfseries 72} (2016) 11}.

\bibitem{hillas}
A.M.~{Hillas}, \emph{{Cerenkov Light Images of EAS Produced by Primary Gamma
  Rays and by Nuclei}},  in \emph{19th International Cosmic Ray Conference
  (ICRC19), Volume 3}, vol.~3 of \emph{International Cosmic Ray Conference},
  p.~445, Aug., 1985.

\bibitem{li-ma}
T.P.~Li and Y.Q.~Ma, \emph{{Analysis methods for results in gamma-ray
  astronomy}}, \href{https://doi.org/10.1086/161295}{\emph{Astrophys. J.}
  {\bfseries 272} (1983) 317}.

\bibitem{lessard}
R.~Lessard et~al., \emph{A new analysis method for reconstructing the arrival
  direction of tev gamma rays using a single imaging atmospheric cherenkov
  telescope},
  \href{https://doi.org/https://doi.org/10.1016/S0927-6505(00)00133-X}{\emph{Astroparticle
  Physics} {\bfseries 15} (2001) 1}.

\bibitem{leslie}
L.P.~Taylor, \emph{Design and performance of the prototype schwarzschild-couder
  telescope camera},  in \emph{these proceedings}, PoS (ICRC2021) 748.

\bibitem{deivid}
D.~Ribeiro, \emph{Prototype schwarzschild-couder telescope for the cherenkov
  telescope array: Commissioning the optical system},  in \emph{these
  proceedings}, PoS (ICRC2021) 717.

\end{thebibliography}\endgroup

%% Full authors list (ONLY FOR COLLABORATIONS)
\clearpage
\section*{Full Authors List: \Coll\ Consortium}

%\noindent \textbf{Note comment afterwards:} Collaborations have the possibility to provide an authors list in xml format which will be used while generating the DOI entries making the full authors list searchable in databases like Inspire HEP. For instructions please go to icrc2021.desy.de/proceedings or contact us under icrc2021proc@desy.de.\\
%
\scriptsize
\noindent
C.~B.~Adams$^{1}$,
G.~Ambrosi$^{2}$,
M.~Ambrosio$^{3}$,
C.~Aramo$^{3}$,
P.~I.~Batista$^{4}$,
W.~Benbow$^{5}$,
B.~Bertucci$^{2, 6}$,
E.~Bissaldi$^{7, 8}$,
M.~Bitossi$^{9}$,
A.~Boiano$^{3}$,
C.~Bonavolont\`a$^{3, 10}$,
R.~Bose$^{11}$,
A.~Brill$^{1}$,
A.~M.~Brown$^{12}$,
J.~H.~Buckley$^{11}$,
R.~A.~Cameron$^{13}$,
R.~Canestrari$^{14}$,
M.~Capasso$^{15}$,
M.~Caprai$^{2}$,
C.~E.~Covault$^{16}$,
D.~Depaoli$^{17, 18}$,
L.~Di~Venere$^{7, 8}$,
M.~Errando$^{11}$,
S.~Fegan$^{19}$,
Q.~Feng$^{15}$,
E.~Fiandrini$^{2, 20}$,
A.~Furniss$^{21}$,
A.~Gent$^{22}$,
N.~Giglietto$^{7, 8}$,
F.~Giordano$^{7, 8}$,
E.~Giro$^{23}$,
R.~Halliday$^{24}$,
O.~Hervet$^{25}$,
J.~Holder$^{26}$,
T.~B.~Humensky$^{27}$,
S.~Incardona$^{28, 29}$,
M.~Ionica$^{2}$,
W.~Jin$^{30}$,
D.~Kieda$^{31}$,
F.~Licciulli$^{8}$,
S.~Loporchio$^{7, 8}$,
G.~Marsella$^{28, 29}$,
V.~Masone$^{3}$,
K.~Meagher$^{22, 32}$,
T.~Meures$^{32}$,
B.~A.~W.~Mode$^{32}$,
S.~A.~I.~Mognet$^{33}$,
R.~Mukherjee$^{15}$,
A.~Okumura$^{34}$,
N.~Otte$^{22}$,
F.~R.~Pantaleo$^{7, 8}$,
R.~Paoletti$^{35, 9}$,
G.~Pareschi$^{36}$,
F.~Di~Pierro$^{18}$,
E.~Pueschel$^{4}$,
D.~Ribeiro$^{1}$,
L.~Riitano$^{32}$,
E.~Roache$^{5}$,
J.~Rousselle$^{37}$,
A.~Rugliancich$^{9}$,
M.~Santander$^{30}$,
R.~Shang$^{38}$,
L.~Stiaccini$^{35, 9}$,
H.~Tajima$^{34}$,
L.~P.~Taylor$^{32}$,
L.~Tosti$^{2}$,
G.~Tovmassian$^{39}$,
G.~Tripodo$^{28, 29}$,
V.~Vagelli$^{40, 2}$,
M.~Valentino$^{10, 3}$,
J.~Vandenbroucke$^{32}$,
V.~V.~Vassiliev$^{38}$,
D.~A.~Williams$^{25}$,
A.~Zink$^{41}$,
\\
\noindent
$^{1}$Physics Department, Columbia University, New York, NY 10027, USA.
$^{2}$INFN Sezione di Perugia, 06123 Perugia, Italy.
$^{3}$INFN Sezione di Napoli, 80126 Napoli, Italy.
$^{4}$Deutsches Elektronen-Synchrotron, Platanenallee 6, 15738 Zeuthen, Germany.
$^{5}$Center for Astrophysics | Harvard \& Smithsonian, Cambridge, MA 02138, USA.
$^{6}$Dipartimento di Fisica e Geologia dell'Universit\`a degli Studi di Perugia, 06123 Perugia, Italy.
$^{7}$Dipartimento Interateneo di Fisica dell'Universit\`a e del Politecnico di Bari, 70126 Bari, Italy.
$^{8}$INFN Sezione di Bari, 70125 Bari, Italy.
$^{9}$INFN Sezione di Pisa, 56127 Pisa, Italy.
$^{10}$CNR-ISASI, 80078 Pozzuoli, Italy.
$^{11}$Department of Physics, Washington University, St. Louis, MO 63130, USA.
$^{12}$Dept. of Physics and Centre for Advanced Instrumentation, Durham University, Durham DH1 3LE, United Kingdom.
$^{13}$Kavli Institute for Particle Astrophysics and Cosmology, SLAC National Accelerator Laboratory, Stanford University, Stanford, CA 94025, USA.
$^{14}$INAF IASF Palermo, 90146 Palermo, Italy.
$^{15}$Department of Physics and Astronomy, Barnard College, Columbia University, NY 10027, USA.
$^{16}$Department of Physics, Case Western Reserve University, Cleveland, Ohio 44106, USA.
$^{17}$Dipartimento di Fisica dell'Universit\`a degli Studi di Torino, 10125 Torino, Italy.
$^{18}$INFN Sezione di Torino, 10125 Torino, Italy.
$^{19}$LLR/Ecole Polytechnique, Route de Saclay, 91128 Palaiseau Cedex, France.
$^{20}$Y.
$^{21}$Department of Physics, California State University - East Bay, Hayward, CA 94542, USA.
$^{22}$School of Physics \& Center for Relativistic Astrophysics, Georgia Institute of Technology, Atlanta, GA 30332-0430, USA.
$^{23}$INAF Osservatorio Astronomico di Padova, 35122 Padova, Italy.
$^{24}$Dept. of Physics and Astronomy, Michigan State University, East Lansing, MI 48824, USA.
$^{25}$Santa Cruz Institute for Particle Physics and Department of Physics, University of California, Santa Cruz, CA 95064, USA.
$^{26}$Department of Physics and Astronomy and the Bartol Research Institute, University of Delaware, Newark, DE 19716, USA.
$^{27}$Science Department, SUNY Maritime College, Throggs Neck, NY 10465.
$^{28}$Dipartimento di Fisica e Chimica "E. Segr\`e", Universit\`a degli Studi di Palermo, via delle Scienze, 90128 Palermo, Italy.
$^{29}$INFN Sezione di Catania, 95123 Catania, Italy.
$^{30}$Department of Physics and Astronomy, University of Alabama, Tuscaloosa, AL 35487, USA.
$^{31}$Department of Physics and Astronomy, University of Utah, Salt Lake City, UT 84112, USA.
$^{32}$Department of Physics and Wisconsin IceCube Particle Astrophysics Center, University of Wisconsin, Madison, WI 53706, USA.
$^{33}$Pennsylvania State University, University Park, PA 16802, USA.
$^{34}$Institute for Space--Earth Environmental Research and Kobayashi--Maskawa Institute for the Origin of Particles and the Universe, Nagoya University, Nagoya 464-8601, Japan.
$^{35}$Dipartimento di Scienze Fisiche, della Terra e dell'Ambiente, Universit\`a degli Studi di Siena, 53100 Siena, Italy.
$^{36}$INAF - Osservatorio Astronomico di Brera, 20121 Milano/Merate, Italy.
$^{37}$Subaru Telescope NAOJ, Hilo HI 96720, USA.
$^{38}$Department of Physics and Astronomy, University of California, Los Angeles, CA 90095, USA.
$^{39}$Instituto de Astronom\'ia, Universidad Nacional Aut\'onoma de M\'exico, Ciudad de M\'exico, Mexico.
$^{40}$Agenzia Spaziale Italiana, 00133 Roma, Italy.
$^{41}$Friedrich-Alexander-Universit\"at Erlangen-N\"urnberg, Erlangen Centre for Astroparticle Physics, D 91058 Erlangen, Germany.

\end{document}